\documentclass[5p]{elsarticle}
\usepackage{amsfonts}
\usepackage{amsmath}
\usepackage{amssymb}
\usepackage{graphicx}
\usepackage{color}

\begin{document}

\title{Dynamics of a bistable Mott insulator to superfluid phase transition in
cavity optomechanics}

\author{K. Zhang}
\address{State Key Laboratory of Precision Spectroscopy, Dept. of Physics, East China Normal University, Shanghai 200062, China}

\author{W. Chen}
\author{P. Meystre}
\address{B2 Institute, Dept. of Physics and College of Optical Sciences, The University of Arizona, Tucson, AZ 85721, USA}


\begin{abstract}
We study the dynamics of the many-body state of ultracold bosons trapped in a bistable optical lattice in an optomechanical resonator controlled by a time-dependent input field. We focus on the dynamics of the many-body system following discontinuous jumps of the intracavity field. We identify experimentally realizable parameters for the bistable quantum phase transition
between Mott insulator and superfluid.
\end{abstract}


\maketitle

\section{Introduction}

The coupling of coherent optical systems to micromechanical devices, combined with breakthroughs in nanofabrication and in ultracold science, has opened up an exciting new field of research, cavity optomechanics. Several groups have now demonstrated very significant cooling of the vibrational motion of a
broad range of moving mirrors, from nanoscale cantilevers to LIGO-class mirrors \cite{coolm}, and there is every reason to believe that the quantum mechanical ground state of motion of these systems will soon be achieved. In a parallel development, ultracold gases as well as Bose-Einstein condensates placed inside optical resonators have been shown to behave under appropriate conditions much like moving mirrors \cite{Essliger,Stamp}. Following these developments, cavity optomechanics is rapidly becoming a very active sub-field of fundamental and applied research at the boundary between AMO physics, condensed matter physics, and nanoscience.

Cavity optomechanics presents considerable promise both in opening the way to address fundamental questions related to pushing quantum mechanics toward increasingly macroscopic systems, but also in applications that span a variety of areas from quantum detection to the coherent control of microscopic atomic and molecular systems and/or of nanoscale devices. Specific examples include
nanomechanical cantilevers coupled to a Bose-Einstein condensate \cite{Hansch07} or to dipolar molecules \cite{Singh08}, to a single atom \cite{zollor09}, oscillating mirrors coupled to atomic vapors \cite{Genes08},
magnetized resonator tip to color center of diamond \cite{Lukin}. Cold atoms can probe the state of the nanostructure, or conversely, the optomechanical cavity setup can serve as a diagnostic tool for the quantum state of atoms trapped inside the resonator \cite{Mekhov07,Chen07}. On a more applied side such systems will enable the development of ultrasensitive force sensors and may find applications in quantum information processing technology.

In recent work we studied the many-body state of ultracold bosons in a bistable optical lattice potential in an optomechanical resonator. We considered explicitly the weak-coupling limit where the coupling between the cavity-field and the movable mirror results in a bistable optical lattice potential for the atoms, and showed how such a cavity plus cold-atom system can be engineered so that a superfluid and a Mott insulator phase are bistable ground states for the ultracold atomic gas \cite{Chen09}.

The present paper extends this study to consider the dynamics of the transition between these two states as the optical potential undergoes a bistable loop. A time-dependent variational principle combined with a Gutzwiller ansatz are introduced to calculate the evolution of the quantum many-body state as a super-Gaussian incident light pulse \cite{Liu02} switches the intracavity optical field. We discuss conditions under which this process can proceed adiabatically, and identify experimental parameters that would permit to test our predictions.

Section II reviews our model of a quantum-degenerate Bose gas confined by the standing wave generated in a Fabry-P\'{e}rot cavity with one movable end mirror, and briefly reviews its steady-state properties. Section III introduces an effective potential description of the system and comments on the conditions required for an adiabatic evolution of the atoms. It then turns to the time evolution of the superfluid order parameter during the bistable superfluid to Mott insulator transition. Finally, Section IV is a conclusion and outlook.

\section{Model}

\begin{figure}
\includegraphics[width=8cm]{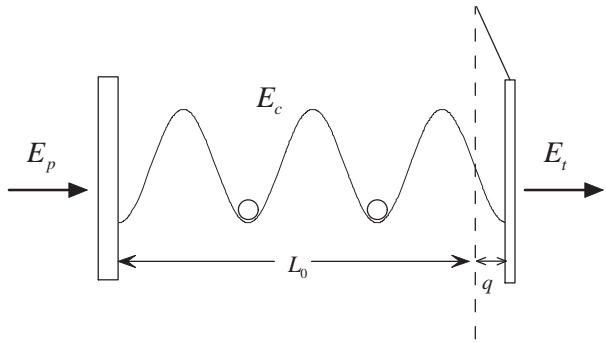}
\caption{ A Fabry-P\'{e}rot cavity with a movable mirror displaced from its equilibrium position by $q$. $E_p$, $E_c$ and $E_t$ are the pump, cavity and transmitted light amplitudes, respectively. }
\label{scheme}
\end{figure}

We consider a Fabry-P\'{e}rot cavity with one fixed end mirror and the other one, of mass $M$, mounted on a spring of frequency $\Omega$ and damping rate $\gamma_m$. An optical field $E_p$ of frequency $\omega _p$ is incident on the fixed mirror. The reflectivity of both mirrors is $R$, with $R \simeq 1$ so that the intracavity field can be approximated by a standing wave of amplitude $E_{c}$. In the absence of that field the movable mirror is at its equilibrium position $q=0$ and the cavity length is $L_0$, see Fig.~\ref{scheme}.

A sample of ultracold bosonic two-level atoms with transition frequency $\omega_a$ is loaded inside the optical lattice formed by the standing wave, and the atoms are assumed to interact with the light field in the weak coupling regime $N g_{0}^{2}/|\Delta_{a}| \ll \kappa_c$, where $N$ is the total number of atoms, $g_{0}$ is the single-photon atom-field coupling strength, $\kappa_c $ is the cavity decay rate, and $\Delta_a=\omega_p- \omega_a$ is the atom-field detuning \cite{Ritsch00}. In this limit the intracavity field has no significant dependence on the atomic distribution and we can investigate the dynamics of the intracavity field by using the theory of an empty cavity. In this paper we describe both the light field and the movable mirror as classical objects, while the center-of-mass motion of the atoms is quantum mechanical. The extension to mirrors cooled near their quantum mechanical ground state of vibration will be considered in future work.

Ignoring the retardation effects due to the propagation of the light field back and forth inside the cavity, the intracavity field is governed by the familiar equation of motion \cite{Ritsch98, Bonif78}
\begin{equation}
\frac{dE_c}{dt}= -( \kappa_c -i\Delta_c ) E_c +\frac{2 \kappa_c}{\sqrt{T}} E_p
\label{lattice}
\end{equation}
where $\Delta_{c}=\omega_p-\omega_c$ is the cavity-pump detuning and $T$ is the transmittivity of the mirrors, with $R+T=1$ in the absence of mirror absorption. The movable mirror is driven by radiation pressure,
\begin{equation}
\frac{d^{2}q}{dt^{2}}+\gamma_m \frac{dq}{dt}+\Omega^{2} q=\frac{F_{\rm{RP}}}{M}
\label{mirror}
\end{equation}
where $F_{\rm RP}\simeq A \epsilon_0 |E_c|^2/4$ is the radiation pressure force, with $A$ being the cross-sectional area of the laser beam.

Equations~\eqref{lattice} and \eqref{mirror} are coupled through the cavity-pump detuning since the cavity frequency is $q$-dependent,
\begin{equation}
\omega_c= n\frac{c \pi}{L_0+q} \simeq \frac{nc\pi}{L_0}\left (1 - q/L_0 \right ),
\end{equation}
where $n$ is the integer associated with the mode closest to the laser frequency in the single-mode theory considered here. The approximate equality assumes that the mirror displacement due to radiation pressure is small compared to both $L_0$ and to the cavity mode wavelength $\lambda_c=2\pi c/\omega_c$.

It is convenient at this point to introduce dimensionless units and to scale times to $\Omega^{-1}$, lengths to $\lambda_p/2$ and electric field amplitudes to $\sqrt{2E_r/c\alpha \epsilon_0}$, where $E_r$ is the photon recoil energy and $\alpha =3\pi c^2 \gamma_a /(2\omega_a^3 \Delta_a)$, with $\gamma_a$ the natural linewidth of the atomic resonance. Equations~\eqref{lattice} and \eqref{mirror} reduce then to the dimensionless form
\begin{align}
\frac{dX}{d\tau }+\left[ 1-i\frac{2\pi }{T}\left( \delta +\xi \right) \right]
\kappa X& =\frac{2\kappa}{\sqrt{T}}Y
\label{dX} \\
\frac{d^{2}\xi }{d\tau ^{2}}+\gamma \frac{d\xi }{d\tau }+\xi & =\beta |X| ^{2}
\label{dxi}
\end{align}
where $\tau$ is the (dimensionless) time, $X$ and $Y$ the intracavity and incident field amplitudes, $\xi$ mirror's displacement, $\kappa$ and $\gamma$ the intracavity optical field and moving mirror damping rates, and $\delta$ the offset from the cavity resonance in the absence
of radiation. Finally, $\beta =(A E_r)/(\lambda_p M \Omega^2 c \alpha)$.

The quantum-degenerate atomic sample, assumed to be at zero temperature is trapped in the optical lattice along cavity axis, its transverse motion being strongly confined by a strong transverse potential. Its Hamiltonian can be then approximated as the one-dimensional form
\begin{equation*}
\hat{H}=\int dx\hat{\Psi}^{\dagger}(x)\left( -\frac{\hbar ^{2}}{2m}\frac{d^{2}}{dx^{2}}+
V_0 \sin^{2} (k_p x) +g\hat{\Psi}^{\dagger} \hat{\Psi} \right) \hat{\Psi}(x)
\end{equation*}
where $\hat{\Psi}(x)$ is the Schr{\"o}dinger field operator, $g$ is the two-body interaction coefficient, $m$ is the atomic mass and $V_0 = \alpha c \epsilon_0 |E_c|^2/2 = |X|^2 E_r$ is the depth of the optical potential.
In the tight-binding approximation this many-body Hamiltonian can be simplified to a single-band Bose-Hubbard Hamiltonian
\begin{equation}
\hat{H}_{\rm{BH}} =-J\sum_{\langle i,j \rangle}\hat{a}_{i}^{\dagger}\hat{a}_{j}
+\frac{U}{2}\sum_{i}\hat{n}_{i}\left( \hat{n}_{i}-1\right)
\label{bh}
\end{equation}
where $\hat{a}_{i}$ is the bosonic annihilation operator for site $i$, $\hat{n}_i$ is the corresponding number operator, the subscript $\langle i,j\rangle$ labeling nearest neighbor pairs. Finally $J$ is the inter-site tunneling matrix element and $U$ is the pair interaction energy. The parameters $J$ and $U$ can be evaluated by expanding the field operators on the Wannier basis of the lowest Bloch band, and then evaluating the pertinent integrals \cite{Jaksch}.

\begin{figure}
\includegraphics[width=8cm]{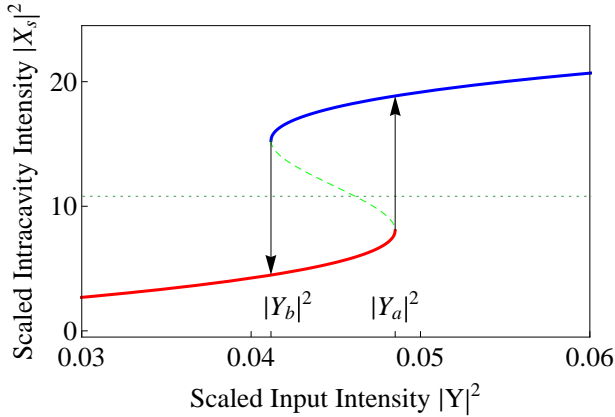}
\caption{Bistable dimensionless intracavity intensity $|X_{s}| ^{2}$ as a function of the dimensionless pump intensity $|Y|^{2}$ for $T=0.01$, initial offset $\delta =-0.0035$ and $\beta =0.0002$. In real units, and for the parameters considered here, the threshold intensities for the bistable domain are 0.74mW and 0.87mW. The steady-state position of the mirror is likewise bistable.}
\label{bisx}
\end{figure}

Setting the time-derivatives to zero yields the steady-state solution
\begin{eqnarray}
X_{s}&=&\frac{2Y/\sqrt{T}}{1-i2\pi \left( \delta +\xi _{s}\right) /T}, \nonumber \\
\xi_{s}&=&\beta | X_{s}|^{2}.
\label{bixi}
\end{eqnarray}
Eliminating the equilibrium displacement $\xi_{s}$ in these equations results in a transcendental equation for the intracavity field intensity $|X_s|^2$,
\begin{equation}
\frac{| X_{s}| ^{2}}{| Y| ^{2}}
=\frac{4/T}{1+4\pi ^{2}\left( \delta +\beta | X_{s}| ^{2}\right)^{2}/T^{2}}.
\label{bissol}
\end{equation}
This equation is known to have a bistable solution for an appropriate choice of parameters, see Fig.~\ref{bisx}. The upper and lower branch with positive slope are generally stable, while the dashed branch with negative slope is unstable \cite{meystre}. It is clear from Eq.~\eqref{bixi} that the steady-state displacement of the movable mirror has the same bistable property as the intracavity field intensity. \footnote{All calculations presented in this paper are for $^{23}$Na and for $\lambda_p$=985nm,  $M$= 0.078g, $\Omega = 2 \pi \times 10$Hz and
mirror reflectivities $R=0.99$.}

As is well known, the ground state of the atomic system is determined by the ratio $J/U$ which can be controlled by varying  the lattice depth $V_0$. For a shallow potential, interwell tunneling  dominates and the many-body ground state is a superfluid, while for a deep enough potential on-site interactions dominate and the atoms enter the Mott-insulator phase with integral filling factor \cite{Jaksch}. The dotted line in Fig.~\ref{bisx} indicates the critical depth $V_0$ at which the superfluid-Mott insulator phase transition occurs, for sodium atoms trapped by a laser of wavelength $985\rm{nm}$. Of particular
interest is the bistable region where the shallow lower branch corresponds to a superfluid phase and the deeper upper branch to a Mott insulator phase. The next section discusses the dynamics of the transition between these two phases as the optical field is varied along the bistable loop.

\section{Dynamics}

Three important time scales are relevant in understanding the dynamics of the system: the characteristic time $\Omega^{-1}$ of the mirror, the cavity buildup time $\kappa ^{-1}$, and the interwell tunneling time $\hbar/J$ of the atoms. In the weak coupling limit the dynamics of the cavity system (the light field plus the movable mirror) is determined by the former two. Furthermore, in the case of a bad cavity the dynamics of the light field is much faster than
that of the movable mirror, $\kappa^{-1} \ll \Omega^{-1}$, so that it can adiabatically follow the instantaneous displacement of the movable mirror. Note that this approximation neglects the non-adiabatic effects that lead to optical damping, see Ref.~\cite{Kipp07} and references therein.

This paper treats the motion of the moving mirror classically as it is not concerned with the dynamics of mirror cooling. We therefore ignore optical damping in the following, and adiabatically eliminate the optical field in the Eq.~\eqref{dxi}, resulting in the nonlinear oscillator equation
\begin{equation}
\frac{d^{2}\xi }{d\tau ^{2}}+\gamma \frac{d\xi }{d\tau }+\xi
=\beta \frac{4|Y|^{2}/T}{1+4\pi ^{2}\left( \delta +\xi \right)^{2}/T^{2}}.
\label{mirr}
\end{equation}

\begin{figure}
\includegraphics[width=9cm]{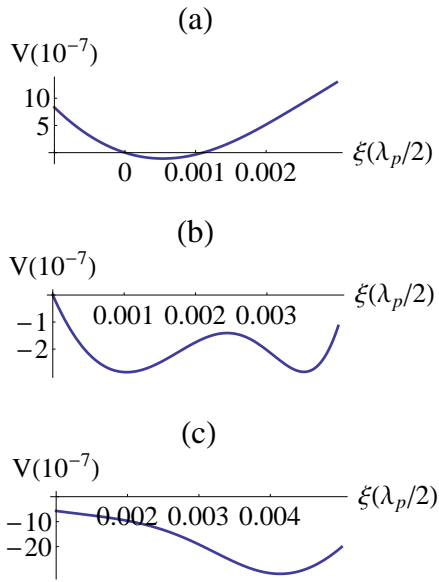}
\caption{Potential $V(\xi)$ for (a) input intensity $|Y|^{2}=0.03$; (b) $|Y|^{2}=0.044$; (c) $|Y|^{2}=0.06$. Other parameters are the same as in Fig.~2. All variables are dimensionless.}
\label{poten}
\end{figure}

To further analyze the dynamics of this nonlinear oscillator we introduce the effective potential \cite{mystreJOSAB}
\begin{equation}
V\left( \xi \right) =\int_{\xi _{0}}^{\xi }\xi ^{\prime }d\xi ^{\prime
}-\int_{\xi _{0}}^{\xi }\beta \frac{4| Y| ^{2}/T}{1+4\pi
^{2}\left( \delta +\xi ^{\prime }\right) ^{2}/T^{2}}d\xi ^{\prime }
\end{equation}
where the first term corresponds to the restoring force on the movable mirror and the second to the radiation pressure force. Figure~\ref{poten} shows that potential for three values of the input intensity. For a weak input field, Fig.~\ref{poten}(a) the potential has a single minimum, but increasing it sees the appearance of a second local minimum, Fig.~\ref{poten}(b), indicative of bistability. Further increasing the input intensity past the bistable region the initial minimum disappears as expected, see Fig.~\ref{poten}(c).

\begin{figure}
\includegraphics[width=9cm]{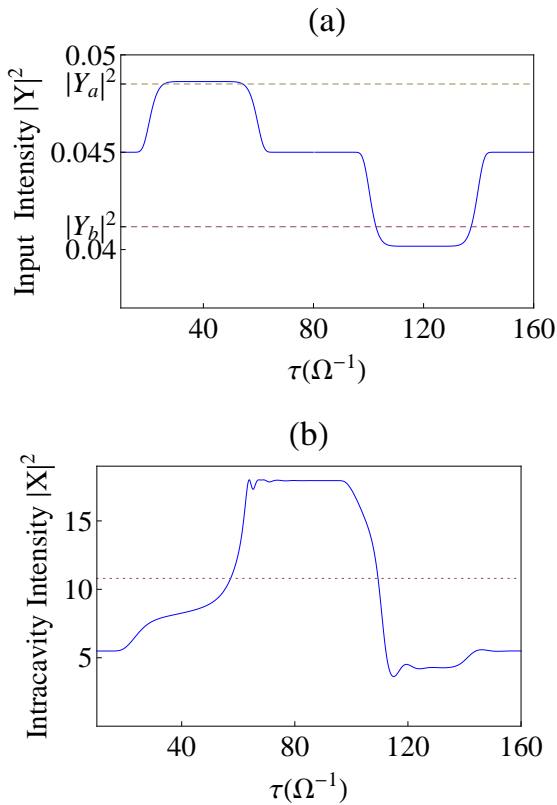}
\caption{ (a) Input intensity $|Y| ^{2}$ as a function of dimensionless time $\tau $. The two dashed lines label the critical intensities at the beginning and the end of the bistable region. (b) Evolution of the intracavity intensity $|X|^2$ controlled by $|Y|^2$. Here $\gamma =0.5$ and the other parameters are as in Fig.~2. The dotted line labels the critical depth of the atomic phase transition.}
\label{dynam}
\end{figure}

For concreteness we assume in the following that the optical field incident on the Fabry-P{\'e}rot is a super-Gaussian pulse, Fig.~\ref{dynam}(a). Since the intracavity field is assumed to follow the motion of the mirror adiabatically, this readily yields the time-dependent optical potential
\begin{equation}
|X(\tau)|^2=\frac{4 |Y(\tau)|^2 /T}{1+4 \pi^2 (\delta +\xi(\tau))^2 / T^2}
\end{equation}

Figure~\ref{dynam}(b) illustrates the dynamics of the intracavity intensity as the incident intensity is varied to switch the system from the lower to the upper bistability branch and back. In this example $|Y(\tau)|^2$ is initially switched to a value only slightly above the critical value $|Y_a|^2$ where the first local minimum of the effective potential degenerates into a plateau, hence it takes a relatively long time to switch to its new steady-state value. Under appropriate circumstances this critical slowing down \cite{Bonif79} can be exploited to allow the atoms to adiabatically follow the potential changes in the superfluid region, see below. Following that slow evolution stage, $|X(\tau)|^2$ grows nearly exponentially as the mirror falls into the newly
formed potential well and completes its transition to the upper branch. For the switch back to the lower branch shown in this example, the lower value of $|Y(\tau)|^2$ is chosen to be significantly below $|Y_b|^2$, so as to reduce the effects of critical slowing down. As expected, $|X(\tau)|^2$ then switches much faster to the lower branch.

The time dependence of the potential $|X(\tau)|^2$ results in Wannier functions that are likewise time-dependent, and hence also in time-dependent tunneling rate $J(\tau)$ and on-site interaction $U(\tau)$. In order for the many-body ground state of the atoms to follow adiabatically the changes in potential, two requirements must be fulfilled.  The first one is that the variation of the potential depth must be slow enough to prevent the occurrence of inter-band excitations, that is the atoms must remain in the first Bloch band at all times. This condition is usually easy to satisfy for atoms with quasi-momentum $q\simeq 0$ because of the existence of a band gap \cite{Phillips02, Dahan96}. The explicit adiabaticity criterion is
$$
|\frac{d}{dt}V_0/E_r|\ll 16 \omega_r.
$$
We note that for the experimental parameters considered here this condition is satisfied for even the fastest transients in Fig.~4(b): The maximum value reached by $|\frac{d}{dt}V_0/E_r|$ is $2 \, \Omega$ which is much less than $16 \, \omega_r$. We therefore assume in the following that that inter-band excitations are negligible, and evaluate $J(t)$ and $U(t)$ with Wannier functions of the first band.

In addition to this single-particle adiabaticity condition, we also need to consider the time scale associated with many-body effects. As the potential depth varies in time, the atoms need a sufficient amount of time to tunnel and redistribute themselves across the lattice, and hence to settle in their new ground state --- think specifically of a transition between a Mott insulator and a superfluid. Although this is not a significant consideration if the atoms remain in the Mott insulator phase, which is quite insensitive to inter-well rearrangement, it is more important in the superfluid region, due to the variation of the tunneling rate $J(t)$\cite{Gericke,Greiner02}. That is the reason why the critical slowing down can be useful in controlling the variation of the potential in the superfluid region. Whether or not the second adiabaticity condition is satisfied can be determined from the time-dependent order parameter which we calculate next.

The evolution of the atomic ground state can be obtained from the time-dependent variational principle \cite{Jaksch02, Damski03},
\begin{equation}
\delta \langle G| i\hbar \frac{\partial }{\partial t}-\hat{H}_{\rm{BH}}(t) |G \rangle =0 ,
\label{tdvp}
\end{equation}
where the Bose-Hubbard Hamiltonian $\hat{H}_{\rm{BH}}(t)$ depends on time via the coefficients $J(t)$ and $U(t)$, and $|G \rangle$ is taken to be given by the time-dependent mean-field Gutzwiller ansatz
\begin{equation}
|G \rangle = \prod_{i}^{N_l} \left( \sum_{n=0}^{\infty}f_n^{(i)}(t) |n \rangle \right).
\end{equation}
Here $N_l$ is the number of lattice sites, $|n \rangle$ are Fock states, and
$f_n^{(i)}(t)$ are probability amplitudes that preserves the normalization of the wave function. We assume that the lattice is uniform with unit filling factor and insert the Gutzwiller ansatz into Eq.~\eqref{tdvp}. This yields the set of coupled nonlinear equations for the amplitudes $f_n(t)$
\begin{eqnarray}
i\hbar \frac{\partial }{\partial t}f_{n} &=& -2J\left( t\right) \left( \sqrt{n%
}f_{n-1}\left\langle a\right\rangle +\left\langle a^{\dagger }\right\rangle
\sqrt{n+1}f_{n+1}\right)   \notag \\
&+&\frac{U\left( t\right) }{2}n\left( n+1\right) f_n
\label{fn}
\end{eqnarray}
where $\langle a\rangle=\sum_{n=0}^{\infty }\sqrt{n+1}f_{n}^{\ast }f_{n+1}$ is the atomic superfluid order parameter. Our numerical results are for an initial optical potential depth of 5$E_r$, for which the tight binding approximation
is valid and the many-body ground state is superfluid.
\begin{figure}
\includegraphics[width=10cm]{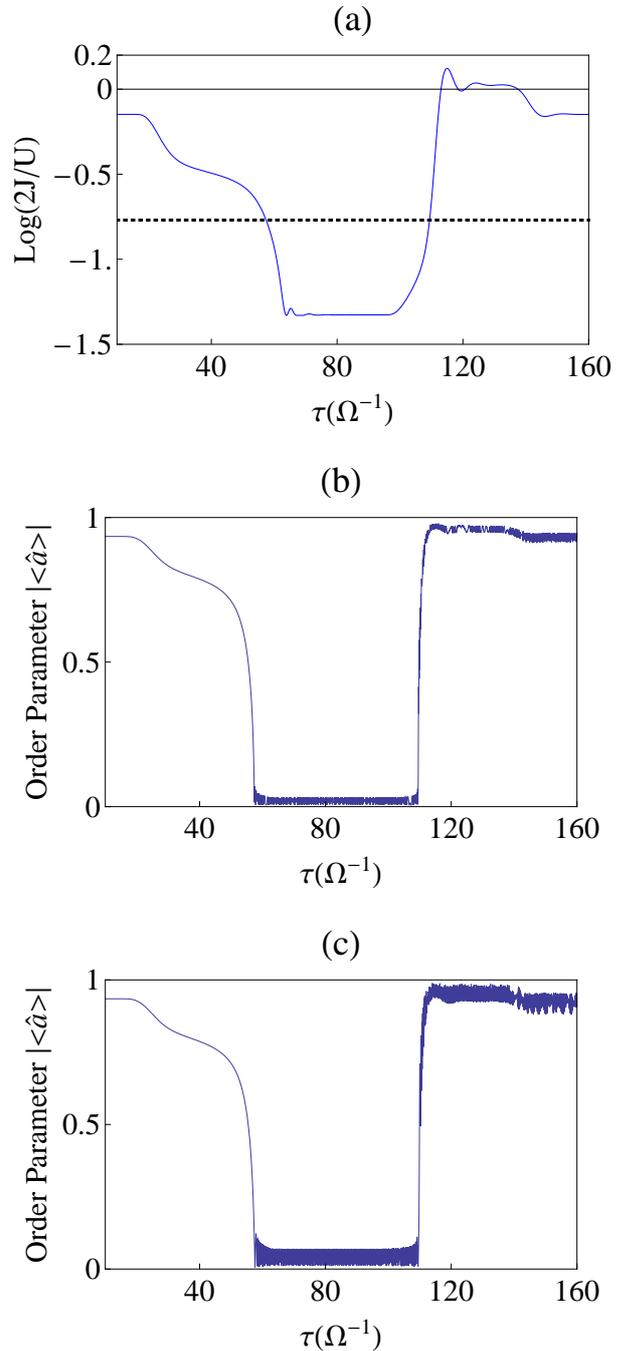}
\caption{ (a) $2J(t)/U(t)$ as a function of the dimensionless time $\tau$. The dotted line indicates the critical value $2J/U=0.17$, corresponding to the potential depth $10.8E_{r}$ for our parameters, for the superfluid and to Mott insulator phase transition. (b) Dynamics of the superfluid order parameter $|\langle a\rangle|$ during the transition from the superfluid to the Mott insulator and back to the superfluid. Here $\lambda_p=985$\,nm
and $\omega_r=E_r/\hbar =2\pi \times 8.9$\,kHz for $^{23}$Na. The mass and vibration frequency of the mirror are $M=0.078$\,g and $\Omega = 2\pi \times 10$\,Hz, respectively. In real units, the full cycle through the bistable loop takes $2.5$\,s. (c) Same as (b) but for $M=0.031$\,g and $\Omega = 2\pi \times 50$\,Hz}
\label{bisphase}
\end{figure}
Figure~\ref{bisphase}(a) shows the ratio $2J(t)/U(t)$ as a function of time. The transition between the superfluid and Mott insulator phases is clearly shown by the absolute value of the superfluid order parameter
$| \langle a\rangle |$ plotted in Fig.~\ref{bisphase}(b). As the ratio $2J/U$ decreases below the critical value $0.17$, corresponding to the critical potential depth $V_0=10.8 E_r$ for sodium atoms, the order parameter rapidly drops to zero, indicating that the atoms enter the Mott insulator phase. When $2J/U$ rises back above that critical value, the order parameter recovers a non-zero value, indicative of the return of the atoms to the superfluid phase.

The small oscillations of the order parameter in the Mott-insulator region are due to collapses and revivals of the condensate with period $T=h/U$ \cite{greiner022, Penna04}. The oscillations following the return of the system to the superfluid region were first predicted in Ref.~\cite{Ehud}. They result from a rapid transition of the system from being a Mott insulator to a superfluid, and can be suppressed via a slower variation of potential depth. The small amplitude of these oscillations, less than 5 percent of the initial value of the order parameter, indicates that the second adiabatic condition (associated with tunneling) is well satisfied in this example.

For comparison Fig.~5(c) shows the time-dependent superfluid order parameter
for a situation here non-adiabatic effects are more apparent. In this case the mass and vibration frequency of the mirror have been changed to 0.031g and $2 \pi \times 50 $Hz.  Since these parameters give the same steady-state solutions as Fig.~2, the form of the evolution of the intracavity intensity is unchanged, but it occurs now on a shorter time scale. The cycle through the bistable loop takes now 0.5s in real units, five times faster than the case of Fig.~5(b). Although the first adiabatic condition is still satisfied for these parameters, it is apparent from Fig.~5(c) that the amplitude of the oscillations in both the Mott insulator and the superfluid regions is much larger than in Fig.~5(b) an indication of the breakdown of  the second adiabatic condition.

We conclude by remarking that the finite temperature of the classical mirror leads to thermal position fluctuations $\delta \xi$ which result in turn in fluctuations for the intracavity field strength the lattice potential depth $\delta V_0$. In order to prevent this effect we request that $\delta V_0 \ll V_{\rm min}$, where $V_{\rm min}$ is the minimum value of the potential depth during the switching process. This inequality gives a mirror temperature limit of $T \ll 10^3 {\rm K}$.

\section{Conclusion}

We have analyzed the dynamics of a quantum-degenerate ultracold sample of bosonic atoms trapped in the optical lattice of a bistable optomechanical system. The variation of intracavity optical lattice depth results in a bistable quantum phase transition between a superfluid and a Mott-insulator ground state. We considered the concrete case of a super-Gaussian incident light pulse to drive the atomic system around the bistable loop, and found that the critical slowing down of the optomechanical system can be useful to prevent the excitation of the atoms in a potential ramping-up stage. Our numerical results indicate that the bistable superfluid to Mott-insulator phase transition should be realizable for realistic experimental parameters. Future work will extend these studies to the strong-coupling regime, where the optical field depends on the motion of the cold atoms as well as the mirror. We will also investigate to which extent the mirror can serve as a measurement device to observe the dynamics of a quantum phase transition nondestructively, in particular in cases where it is cooled to near its quantum mechanical ground state of vibration, Additional goals include the study of the optically induced quantum correlations and entanglement between the ultracold atomic system and the mirror, the quantum control of the mirror motion by the atoms, and conversely of the atoms by the nanomechanical system, and finally the development of novel types of quantum sensors.

\section*{Acknowledgements}

This paper is dedicated to the memory of Krzysztof W{\' o}dkiewicz, excellent physicist and even better friend, with whom it has been great fun to push the frontiers of quantum optics over many years and in many places, from the old Max Planck Institute for Quantum Optics "theory container" to the American Southwest.

We thank M. Bhattacharya and D. Goldbaum for useful discussions. This work is supported in part by the US Office of Naval Research, by the National Science Foundation, and by the US Army Research Office.

\end{document}